\documentstyle[preprint,aps]{revtex}

\tighten
\begin{document}

\title{
       Shell-model calculations for the three-nucleon system 
}
\medskip

\author{
        P. Navr\'atil\footnote{On leave of absence from the
   Institute of Nuclear Physics,
                   Academy of Sciences of the Czech Republic,
                   250 68 \v{R}e\v{z} near Prague,
                     Czech Republic.}
     and B. R. Barrett
        }

\medskip

\address{
                   Department of Physics,
                   University of Arizona,
                   Tucson, Arizona 85721
}

\maketitle

\bigskip

\begin{abstract}
We use Faddeev's decomposition to solve the shell-model problem for 
three nucleons. The dependence on harmonic-oscillator excitations
allowed in the model space, up to $32 \hbar\Omega$ in the present 
calculations, and on the harmonic-oscillator frequency is studied. 
Effective interactions derived from Nijmegen II and Reid93 
potentials are used in the calculations. 
The binding energies obtained are close to those calculated by other
methods. The structure of the Faddeev equations is discussed and 
a simple formula for matrix elements of the permutation operators 
in a harmonic-oscillator basis is given. The Pauli principle is properly 
treated in the calculations.  
\end{abstract}

\bigskip
\bigskip
\bigskip

\narrowtext



\section{Introduction}
\label{sec1}

Many different methods have been used to solve the 
three-nucleon problem in the past. The most viable 
approach appears to be the Faddeev method \cite{Fad60}.
It has been successfully applied to solve the three-nucleon bound-state
problem for various nucleon-nucleon 
potentials \cite{PFGA80,PFG80,CPFG85,FPSS93}.
The most complex calculations of this kind include
up to 34 channels, when all the $j\leq 4$ waves are taken
into account. The precision achieved in these calculations 
is better than 1\% \cite{CPFG85,FPSS93}.

On the other hand, when studying the properties of more
complex nuclei one typically resorts to the shell model.
In that approach, the harmonic-oscillator basis is used in a truncated 
model space. Instead of the free nucleon-nucleon potential, 
one uses effective interactions inside the model space. 
Examples of such calculations are the
large-basis no-core shell-model calculations that have recently been 
performed \cite{ZVB,ZBVHS,NB96}. 
In these calculations all nucleons are active,
which simplifies the effective interaction as no hole states
are present. The effective interaction is
determined for a system of two nucleons in a harmonic-oscillator well
interacting by the nucleon-nucleon potential and is subsequently
used in the many-particle calculations. 

In the present paper we combine the application of the shell-model
approach to the three-nucleon system with the Faddeev method.
We use Faddeev's decomposition for the basis states and remove the
center-of-mass term. This leads to a significant simplification of the 
problem and allows us to extend considerably the model space.
We can study the convergence properties of the results with the
increasing model space. If convergence is achieved, then this
approach leads to the exact solution of the three-nucleon problem
and is, thus, complementary to the traditional calculations, based
on either the differential equation solutions in the configuration
space or integral equation solitions in the momentum space treatments.  

In addition to the attempt of solving the three-nucleon problem
exactly, the present method serves primarily as a test of 
the shell-model approach. In particular, it allows us to test effective 
interactions used in standard shell-model applications.
The present approach has several advantages. First,
any number of partial waves can be included. 
Second, the calculation is simplified by using a compact formula
for the matrix elements of the permutation operators
in the harmonic-oscillator basis.
Also, because of the way we do the model-space truncation, we keep
equivalence of the Faddeev and Schr\"odinger equations throughout 
the calculation.

In section \ref{sec2} we discuss the shell-model Hamiltonian
with a bound center-of-mass, the Faddeev equation,  
and the methods used to derive the starting-energy-independent
effective interaction. 
Results of the calculations for the three-nucleon system are presented in
section \ref{sec3}. In particular, we discuss the 
harmonic-oscillator frequency and the model-space-size
dependences. 
Conclusions are given in section \ref{sec4}.

\section{Shell-model hamiltonian and the Faddeev approach}
\label{sec2}

In most shell-model studies the one- plus two-body Hamiltonian
for the A-nucleon system, i.e.,
\begin{equation}\label{ham}
H=\sum_{i=1}^A \frac{\vec{p}_i^2}{2m}+\sum_{i<j}^A 
V_{\rm N}(\vec{r}_i-\vec{r}_j) \; ,
\end{equation}
where $m$ is the nucleon mass and $V_{\rm N}(\vec{r}_i-\vec{r}_j)$ 
the nucleon-nucleon interaction,
is modified by adding the center-of-mass harmonic-oscillator potential
$\frac{1}{2}Am\Omega^2 \vec{R}^2$, 
$\vec{R}=\frac{1}{A}\sum_{i=1}^{A}\vec{r}_i$.
This potential does not influence intrinsic properties of the 
many-body system. It provides, however, a mean field felt by each nucleon
and allows us to work with a convenient harmonic-oscillator basis.
The modified Hamiltonian, depending on the harmonic-oscillator 
frequency $\Omega$, can be cast into the form
\begin{equation}\label{hamomega}
H^\Omega=\sum_{i=1}^A \left[ \frac{\vec{p}_i^2}{2m}
+\frac{1}{2}m\Omega^2 \vec{r}^2_i
\right] + \sum_{i<j}^A \left[ V_{\rm N}(\vec{r}_i-\vec{r}_j)
-\frac{m\Omega^2}{2A}
(\vec{r}_i-\vec{r}_j)^2
\right] \; .
\end{equation}
The one-body term of the Hamiltonian (\ref{hamomega}) can be re-written
as a sum of the center-of-mass term
$H^\Omega_{\rm cm}=\frac{\vec{P}_{\rm cm}^2}{2Am}
+\frac{1}{2}Am\Omega^2 \vec{R}^2$,
$\vec{P}_{\rm cm}=\sum_{i=1}^A \vec{p}_i$,
and a term depending on relative coordinates only.
In the present application we use a basis, which explicitly separates
center-of-mass and relative-coordinate wave functions. Therefore,
the center-of-mass term contribution is trivial and will 
be omitted from now on. 

For a three-nucleon system, i.e., $A=3$, 
the following transformation of the coordinates 
\begin{mathletters}\label{rtrans}
\begin{eqnarray}
\vec{r}&=&\sqrt{\textstyle{\frac{1}{2}}}(\vec{r}_1-\vec{r}_2) \; ,
\\
\vec{y}&=&\sqrt{\textstyle{\frac{2}{3}}}[\textstyle{\frac{1}{2}}(\vec{r}_1
+\vec{r}_2)-\vec{r}_3] \; ,
\end{eqnarray}
\end{mathletters}
and, similarly, of the momenta
\begin{mathletters}\label{ptrans}
\begin{eqnarray}
\vec{p}&=&\sqrt{\textstyle{\frac{1}{2}}}(\vec{p}_1-\vec{p}_2) \; ,
\\
\vec{q}&=&\sqrt{\textstyle{\frac{1}{6}}}(\vec{p}_1
+\vec{p}_2)-\sqrt{\textstyle{\frac{2}{3}}}\vec{p}_3 \; ,
\end{eqnarray}
\end{mathletters}
can be introduced that brings the relative-coordinate part of the
one-body harmonic-oscillator Hamiltonian into the form
\begin{equation}\label{H0}
H_0 =  \frac{\vec{p}^2}{2m} + \frac{1}{2}m\Omega^2 \vec{r}^2
     + \frac{\vec{q}^2}{2m} + \frac{1}{2}m\Omega^2 \vec{y}^2 \; .
\end{equation}
Eigenstates of this Hamiltonian, 
\begin{equation}\label{hobas}
|n l s j t, {\cal N} {\cal L} {\cal S} {\cal J} \tau, J T\rangle \; ,
\end{equation}
are then used as the basis for our calculation.
Here $n, l$ and ${\cal N}, {\cal L}$ are the harmonic-oscillator quantum numbers
corresponding to the harmonic oscillators associated with the coordinates and 
momenta $\vec{r}, \vec{p}$ and $\vec{y}, \vec{q}$, respectively. 
The quantum numbers $s,t,j$ describe the spin, isospin and angular momentum
of the relative-coordinate partial channel of particles 1 and 2. 
${\cal S}=\textstyle{\frac{1}{2}}$
and $\tau=\textstyle{\frac{1}{2}}$ are the spin and  isospin of the third particle,
while ${\cal J}$ is the angular momentum of the third particle relative to the
center of mass of particles 1 and 2. The $J$ and $T$ are the total angular 
momentum and the total isospin, respectively. Note that for the $^3$H nucleus 
$J=\textstyle{\frac{1}{2}}$ and $T=\textstyle{\frac{1}{2}}$.

The Faddeev equation for the bound sytem can be written in the form
\begin{equation}\label{Fadeq}
\tilde{H}|\phi\rangle = E|\phi\rangle \; ,
\end{equation}
with
\begin{equation}\label{Fadham}
\tilde{H}= H_0 + V(\vec{r}) {\cal T} \; .
\end{equation}
Here, 
$V(\vec{r})=V_{\rm N}(\sqrt{2}\vec{r})-\textstyle{\frac{1}{3}}m
\Omega^2 \vec{r}^2$
is the potential and ${\cal T}$, which has the properties of 
a metric operator \cite{PFG80,SGH}, is given by
\begin{equation}\label{metric}
{\cal T}=1+{\cal T}^{(-)}+{\cal T}^{(+)} \; ,
\end{equation}
with ${\cal T}^{(+)}$ and ${\cal T}^{(-)}$ the cyclic and the anticyclic 
permutation 
operators, respectively. We derived a simple formula for the matrix elements of 
${\cal T}^{(-)}+{\cal T}^{(+)}$ in the basis (\ref{hobas}), namely
\begin{eqnarray}\label{t13t23}
&&\langle n_1 l_1 s_1 j_1 t_1, {\cal N}_1 {\cal L}_1 {\cal S} 
{\cal J}_1 \tau, J T | {\cal T}^{(-)}+{\cal T}^{(+)} |  
n_2 l_2 s_2 j_2 t_2, {\cal N}_2 {\cal L}_2 {\cal S} 
{\cal J}_2 \tau, J T\rangle = - \delta_{N_1,N_2} \nonumber \\
&& \times \sum_{LS} \hat{L}^2 \hat{S}^2
\hat{j}_1 \hat{j}_2 \hat{\cal J}_1 \hat{\cal J}_2 \hat{s}_1 \hat{s}_2
\hat{t}_1 \hat{t}_2 (-1)^L
          \left\{ \begin{array}{ccc} l_1   & s_1   & j_1   \\
          {\cal L}_1  & \textstyle{\frac{1}{2}}  & {\cal J}_1 \\
                                     L   & S  & J
\end{array}\right\}
          \left\{ \begin{array}{ccc} l_2   & s_2   & j_2   \\
       {\cal L}_2  & \textstyle{\frac{1}{2}}   & {\cal J}_2 \\
                                     L   & S  & J
\end{array}\right\}
  \left\{ \begin{array}{ccc} \textstyle{\frac{1}{2}} & \textstyle{\frac{1}{2}} 
               & s_1 \\
             \textstyle{\frac{1}{2}}  &  S  & s_2
\end{array}\right\}
  \left\{ \begin{array}{ccc} \textstyle{\frac{1}{2}} & \textstyle{\frac{1}{2}} 
               & t_1 \\
             \textstyle{\frac{1}{2}}  &  T  & t_2
\end{array}\right\}
\nonumber \\
&&\times \left[(-1)^{s_1+s_2+t_1+t_2-{\cal L}_1-l_1} 
\langle {\cal N}_1 {\cal L}_1 n_1 l_1 L 
| n_2 l_2 {\cal N}_2 {\cal L}_2 L \rangle_{\rm d=3}
+\langle n_1 l_1 {\cal N}_1 {\cal L}_1 L 
| {\cal N}_2 {\cal L}_2 n_2 l_2 L \rangle_{\rm d=3}
\right] \; ,
\end{eqnarray}
where $N_i=2n_i+l_i+2{\cal N}_i+{\cal L}_i, i\equiv 1,2$, 
$\hat{j}=\sqrt{2j+1}$ 
and $\langle {\cal N}_1 {\cal L}_1 n_1 l_1 L 
| n_2 l_2 {\cal N}_2 {\cal L}_2 L \rangle_{\rm d=3}$
is the general harmonic-oscillator bracket for two particles with mass 
ratio 3 as defined, e.g.,
in Ref. \cite{Tr72}, where a compact formula is also given for calculating 
the brackets. The expression (\ref{t13t23}) can be derived by examining 
the action
of ${\cal T}^{(+)}$ and ${\cal T}^{(-)}$ on the basis states (\ref{hobas}).
A similar derivation for a different basis is described, e.g., 
in Ref. \cite{HKT72}.

Note that the eigensystem of the metric ${\cal T}$ (\ref{metric}) consists
of two subspaces. The first subspace has eigenstates with the eigenvalue
3, which form totally antisymmetric physical states, while the second 
subspace has eigenstates with
the eigenvalue 0, which form a not completely antisymmetric, unphysical 
subspace of states. It is possible to hermitize the 
Hamiltonian (\ref{Fadham}) on the physical subspace, where it is 
quasi-Hermitian
(see the discussion of quasi-Hermitian operators, e.g., in Ref. \cite{SGH}) .
The Hermitized Hamiltonian takes the form
\begin{equation}\label{Fadhamh}
\bar{H}= H_0 + \bar{{\cal T}}^{1/2}V(\vec{r})\bar{{\cal T}}^{1/2} \; ,
\end{equation}
where $\bar{{\cal T}}$ operates on the physical subspace only.

Apparently, the interaction $V(\vec{r})$ is diagonal in the quantum numbers
${\cal N, L, J}$ (and also in $s,j,t$ due to the properties of the
nucleon-nucleon potential). The metric ${\cal T}$ (\ref{metric}) is, 
on the other hand, diagonal in 
$N=2n+l+2{\cal N}+{\cal L}$. Note that any basis truncation other 
than one of the type
$N\le N_{\rm max}$ violates, in general, the Pauli principle and mixes 
physical and unphysical states. 
Here, $N_{\rm max}$ characterizes the maximum of total allowed 
harmonic-oscillator quanta in the model space and is an input parameter
of the calculation. 
At the same time, the truncation into totally allowed
oscillator quanta $N\le N_{\rm max}$ preserves the equivalence of the
Hamiltonians (\ref{Fadham}) and (\ref{Fadhamh}) on the physical subspace.

From solving two-nucleon systems 
in a harmonic-oscillator well, interacting by soft-core potentials,
one learns that excitations up to about $300\hbar\Omega$ ($N_{\rm max}=300$)
are required to get almost exact solutions. We anticipate,
therefore, that at least the same number of excitations should
be allowed to solve the three-nucleon system
using the formalism discussed above. The Faddeev formulation
has the obvious advantage compared with the traditional shell-model
approach that the center-of-mass coordinate
is explicitly removed. Even then, it is presently not
feasible to solve the eigenvalue problem either for (\ref{Fadham})
or for (\ref{Fadhamh}) in such a large space.
On the other hand, shell-model calculations are always performed by
employing effective interactions tailored to a specific
model space. In practice, these effective interactions can
never be calculated exactly as, in general, for an A-nucleon system 
the A-body effective interaction is required. 
Consequently, large model spaces
are desirable. In that case, the calculation
should be less affected by any imprecision of the effective
interaction. 
The same is true for the evaluation of any observable characterized
by an operator. In the model space, renormalized effective operators 
are required. The larger the model space, the less renormalization
is needed.  
We may take advantage of the present approach
to perform shell-model calculations in significantly larger
model spaces than are possible in conventional shell-model
approach, particularly
when using a Hermitized Hamiltonian (\ref{Fadhamh}).
At the same time we can investigate convergence properties
of effective interactions. If convergence is achieved,
we should obtain the exact solution, since 
we recover the original full-space problem as 
$N_{\rm max}\rightarrow\infty$,
provided that the condition $V_{\rm eff}\rightarrow V$ is satisfied
in this limit.

Usually, the effective interaction is approximated by a 
two-body effective interaction determined from a two-nucleon
system.    
In the present calculations we replace matrix elements of the potential
$V(\vec{r})$ by matrix elements of an effective two-body
interaction, derived in a straightforward manner for each
relative-coordinate partial channel. 
The relevant two-nucleon Hamiltonian is then
\begin{equation}\label{hamomega2}
H_2\equiv H_{02}+V=
\frac{\vec{p}^2}{2m}
+\frac{1}{2}m\Omega^2 \vec{r}^2
+ V_{\rm N}(\sqrt{2}\vec{r})-\frac{m\Omega^2}{3}\vec{r}^2 \; ,
\end{equation}
which can be solved as a differential equation or, alternatively,
can be diagonalized in a sufficiently large harmonic oscillator
basis. The latter possibility is, obviously, not applicable 
for hard-core nucleon-nucleon potentials. 

To construct the effective interaction we employ
the Lee-Suzuki \cite{LS80} similarity transformation
method, which gives the effective interaction in the form
$P V_{\rm eff}P = P V P + PV Q\omega P$,
with $\omega$ the transformation operator satisfying $\omega=Q \omega P$,
and $P$, $Q=1-P$, the projectors on the model and the complementary 
spaces, respectively. 
Our calculations start with exact solutions of the Hamiltonian
(\ref{hamomega2}) and, consequently, we construct
the operator $\omega$ and, then, the effective interaction directly
from these solutions. Let us denote the relative-coordinate two-nucleon 
harmonic-oscillator states, which form the model space, 
as $|\alpha_P\rangle$,
and those which belong to the Q-space, as $|\alpha_Q\rangle$.
Then the Q-space components of the eigenvector $|k\rangle$ of
the Hamiltonian (\ref{hamomega2}) can be expressed as a combination
of the P-space components with the help of the operator $\omega$
\begin{equation}\label{eigomega}  
\langle\alpha_Q|k\rangle=\sum_{\alpha_P}
\langle\alpha_Q|\omega|\alpha_P\rangle \langle\alpha_P|k\rangle \; .
\end{equation}
If the dimension of the model space is $d_P$, we may choose a set
${\cal K}$ of $d_P$ eigenevectors, 
for which the relation (\ref{eigomega}) 
will be satisfied. Under the condition that the $d_P\times d_P$ 
matrix $\langle\alpha_P|k\rangle$ for $|k\rangle\in{\cal K}$
is invertible, the operator $\omega$ can be determined from 
(\ref{eigomega}).  In the present application we select the lowest states
obtained in each channel. Their number is given by the number of basis
states satisfying $2n+l\leq N_{\rm max}$. 
Once the operator $\omega$ is determined the effective hamiltonian
can be constructed as follows 
\begin{equation}\label{effomega}
\langle \gamma_P|H_{2\rm eff}|\alpha_P\rangle =\sum_{k\in{\cal K}}
\left[
\langle\gamma_P|k\rangle E_k\langle k|\alpha_P\rangle
+\sum_{\alpha_Q}\langle\gamma_P|k\rangle E_k\langle k|\alpha_Q\rangle
\langle\alpha_Q |\omega|\alpha_P\rangle\right] \; .
\end{equation}
It should be noted that 
$P|k\rangle=\sum_{\alpha_P}|\alpha_P\rangle\langle\alpha_P|k\rangle$
for $|k\rangle\in{\cal K}$ is a right eigenvector of (\ref{effomega})
with the eigenvalue $E_k$.

This Hamiltonian, when diagonalized in a model-space basis, reproduces
exactly the set ${\cal K}$ of $d_P$ eigenvalues $E_k$. Note that
the effective Hamiltonian is, in general, quasi-Hermitian. 
It can be hermitized by a similarity transformation 
determined from the metric operator $P(1+\omega^\dagger\omega)P$. 
The Hermitian Hamiltonian is then given by \cite{S82SO83}
\begin{equation}\label{hermeffomega}
\bar{H}_{\rm 2eff}
=\left[P(1+\omega^\dagger\omega)P\right]^{1/2}
H_{\rm 2eff}\left[P(1+\omega^\dagger\omega)
P\right]^{-1/2} \; .
\end{equation}

Finally, the two-body effective interaction used 
in the present calculations
is determined from the two-nucleon effective Hamiltonian 
(\ref{hermeffomega}) as $V_{\rm eff}=\bar{H}_{\rm 2eff}-H_{02}$.

\section{Application to the three-nucleon problem}
\label{sec3}

In this section we discuss the results of application of the formalism
outlined in section \ref{sec2} for the $^3$H system.
In the calculations we use the Nijmegen II nucleon-nucleon
potential \cite{SKTS} corrected in the $^1P_1$ wave \cite{Spc},
and the Reid93 nucleon-nucleon potential \cite{SKTS}. 
We work in the isospin formalism; the charge invariant potential 
$V_{\rm N} =\frac{2}{3} V_{nn}+\frac{1}{3} V_{np}$
is used for each $T=1$ wave \cite{FGP87}.

The two-body effective interaction employed in the calculation
is derived from the Eqs. 
(\ref{eigomega})-(\ref{hermeffomega}). 
Our model space is characterized by the condition $N\le N_{\rm max}$,
$N=2n+l+2{\cal N}+{\cal L}$. The condition for the relative-coordinate
effective-interaction model space is then $2n+l\le N_{\rm max}$.
When diagonalizing the two-nucleon relative-coordinate Hamiltonian 
(\ref{hamomega2}) in the full space we truncate the harmonic-oscillator 
basis by keeping only the states with $n\le 152$. 
The error caused by this truncation can be estimated,
as the system can be solved as a differential equation. 
We found that the low-lying eigenvalues
obtained in the two calculations do not differ by more than 
$\approx 10^{-3}$ MeV and in most cases by much less. 
The lowest eigenvalues are typically of the order of 
$10^1$ MeV.
Note that this error decreases with increasing $\Omega$.
We calculated the effective
interactions up to $N_{\rm max}=32$, as required in the present
aplication.

Once the effective interaction is found we may directly diagonalize 
the non-Hermitian Hamiltonian (\ref{Fadham}) in the basis
(\ref{hobas}) truncated by $N_{\rm max}$ and with $V$ replaced by
$V_{\rm eff}$.

On the other hand, a calculation with the hermitized Hamiltonian 
(\ref{Fadhamh}) 
can be performed in three steps. First the effective
interaction is calculated for each relative-coordinate
partial channel. Second the metric ${\cal T}$ (\ref{metric})
is diagonalized for each $N$ up to $N_{\rm max}$. The 
physical eigenvectors corresponding to the eigenvalue 3
are selected and used, finally, as a new basis
in which the Hamiltonian (\ref{Fadhamh}) is diagonalized.
As the number of physical states is about a third
of the number of all original basis states (\ref{hobas}),
it is more efficient to diagonalize the Hamiltonian
(\ref{Fadhamh}) than the non-Hermitian Hamiltonian
(\ref{Fadham}), in particular for higher values
of $N_{\rm max}$. In fact, for $N_{\rm max}>22$
we used the Hermitian Hamiltonian only.
Note that the Hamiltonians (\ref{Fadham}) and (\ref{Fadhamh})
have identical spectra of the physical states, provided that 
no other truncation than $N\leq N_{\rm max}$ is allowed.
The unphysical eigenstates of $\tilde{H}$ (\ref{Fadham})
have energies corresponding to the unperturbed harmonic 
oscillator, starting at $3\hbar\Omega$.

In Figs. \ref{figniie}-\ref{figr93r} we present the results for 
the ground-state energies and point-nucleon radii, calculated from
$\langle r^2\rangle = \frac{1}{A} \sum_{i=1}^A 
\langle (\vec{r}_i-\vec{R})^2\rangle $,
obtained with the Nijmegen II and the Reid93 nucleon-nucleon 
potentials, respectively. Our calculation starts at
$N_{\rm max}=8$, which corresponds to a model space easily 
accessible with the traditional shell-model calculations.
In Ref. \cite{NB96} we performed an $8\hbar\Omega$ 
calculation for $^3$H using a slightly different effective
interaction than we employ here but derived in an analogous
way. Note that it is straightforward to transform the 
relative-coordinate effective interaction used in the present 
calculations to the two-particle basis used for the shell-model
input by the standard transformation \cite{HG83}. We used
the transformed interaction in the $8\hbar\Omega$ space to test
our results.
The shell-model diagonalization was performed by using the
Many-Fermion-Dynamics Shell-Model Code \cite{VZ94} and
we obtain the same answers from both the present calculation and
the shell-model calculation. The present calculation has,
obviously, much smaller dimension.

As the results depend on $N_{\rm max}$ and $\Omega$
introduced in Eq. (\ref{hamomega}), we must test the convergence
with regard to both of these parameters.
With increasing $N_{\rm max}$ the calculations grow tedious.
We performed the calculations up to $N_{\rm max}=32$ for a wide range
of the harmonic-oscillator frequencies $\Omega$ with values
typical for standard shell-model calculations varying 
from $\hbar\Omega=14$ MeV to $\hbar\Omega=24$ MeV.
In general, we observe a slow convergence with increasing $N_{\rm max}$.
An unusual feature is the convergence from below. This is caused
by the use of effective interactions instead of the
free nucleon-nucleon interaction. The effective interactions
we employ are too strong.
We have not reached the convergence with respect to $\Omega$
in the whole range studied. However, for the values 
$\hbar\Omega=22-24$ MeV our results almost reach convergence with
$N_{\rm max}=32$, 
in particular for the Reid93 potential.
We note that the traditional 34-channel Faddeev calculation,
as reported in Ref. \cite{FPSS93}, gives the binding energies
7.62 MeV and 7.63 MeV for the Nijmegen II and the Reid93 
nucleon-nucleon potentials, respectively. We present these values
in Figs. \ref{figniie}, and \ref{figr93e} as dotted lines for
comparison. From the figures it is apparent that we are obtaining 
virtually the same values in the calculations which start to converge.   
When comparing the results for the two different potentials used 
we can see larger sensitivity to $\Omega$ of the Reid93 calculation.
On the other hand, the calculation using the Nijmegen II potential
is slower in reaching the stability with respect to $N_{\rm max}$.
While the ground-state energy calculation begins to stabilize for the
largest values of $\Omega$ employed, the radius calculation has not 
reached complete stability for any of the $\Omega$ values within 
the model spaces we used. 

As a further test on the stability and convergence of the method 
we analyzed the ground-state wave functions and calculated 
the probability of $S, P, D, S', S''$ states. In Fig. \ref{figwave} 
we show the $D$-state and $S'$-state
probabilities as a function of $N_{\rm max}$ for the Reid93 calculations 
with $\hbar\Omega=19$ and 24 MeV. We observe a good stability 
with respect to
$N_{\rm max}$ and little dependence on $\Omega$ for larger model spaces. 
The $D$-state probability approaches $8.4\%$, $S'$-state probability 
percentage reaches $1.2\%$. Not shown in the figure are the calculated 
$P$-state and $S''$-state percentage probabilities, for which we get
$0.06\%$, and $\approx 10^{-5}\%$, respectively. The present 
numbers are in agreement with those obtained using other nucleon-nucleon
potentials \cite{CPFG85}. The $D$-state percentage is approximately
1.5 times the corresponding $D$-state percentage of deuteron ($5.7\%$).

In addition to the calculations discussed so far, we also 
computed properties of $^3$He, with the focus on obtaining 
the binding-energy difference between $^3$H and $^3$He. 
In those calculations the Coulomb potential was added to
the proton-proton potential and the averaged potential
$V_{\rm N} =\frac{2}{3} V_{pp}+\frac{1}{3} V_{np}$
was eventually used for each $T=1$ wave \cite{FGP87}. The binding-energy 
differences obtained using the Reid93 potential and $\hbar\Omega=19$ 
and 24 MeV are presented in Fig. \ref{figcoul}. For larger model spaces 
we get an almost $\Omega$-independent difference. The binding-energy 
splitting shows convergence with increasing $N_{\rm max}$. It decreaces
with $N_{\rm max}$ in correlation with increasing 
point-nucleon radius and approaches 0.66 MeV. This result is again 
in agreement with those obtained 
using other nucleon-nucleon potentials \cite{CPFG85}.
Note that the experimental value of the binding-energy 
difference is 0.764 MeV. To test the quality of the approximation used 
for the potential averaging and limitation to $T=\frac{1}{2}$, 
we performed an $N_{\rm max}=8$ calculation
with complete isopin breaking using proton-neutron formalism. 
The present method can be used to perform calculations 
with isospin breaking. For this particular calculation, however, 
we employed the Many-Fermion-Dynamics Shell-Model Code. 
The effective interaction was 
calculated separately for the proton-proton and proton-neutron systems,
respectively, and transformed to the two-particle basis as discussed
earlier in this Section. In this way we found that the binding energy 
obtained with and without isospin breaking differs by 11 keV 
and the nucleon radius differs 
by less than 0.001 fm in a calculation with $\hbar\Omega=19$ MeV.
This confirms, that the limitation to $T=\frac{1}{2}$ together with 
the potential averaging provides an excellent approximation.    

We stressed before that no other basis truncation than 
$N\leq N_{\rm max}$
was used. That means that we keep all the relative-coordinate
channels in the basis. In most calculations, however, 
we set the nucleon-nucleon potential $V_{\rm N}$ to zero for $j>6$. 
We also performed calculations with $V_{\rm N}$ set to zero for $j>4$. 
The largest contribution of the $j=5,6$ waves to the binding energy 
we observed was about 5 keV. This contribution increases with
$N_{\rm max}$ and $\Omega$. Moreover,
we performed several calculations
with $V_{\rm N}$ non-zero up to $j=9$, and found that the 
ground-state energy is affected by less than 0.3 keV compared
to the $j\leq 6$ calculations. Note that
such a truncation of $V_{\rm N}$ does not imply that $V=0$,
see Eq. (\ref{Fadham}). Also, this type of potential truncation
is not the same as used in the traditional Faddeev calculations
\cite{PFGA80,PFG80,CPFG85,FPSS93}. The difference is, that
there is no truncation in the treatment of ${\cal T}$ (\ref{metric})
in the present calculations.

It should be noted that the calculated binding energy obtained
from calculations employing nucleon-nucleon potentials fitted
to the two-nucleon scattering underbind $^3$H by about 0.8 MeV,
as its experimental binding energy is 8.48 MeV.
Suggested solutions to this problem include the use of three-body
forces, non-local potentials and relativistic corrections \cite{MSS96}.

\section{Conclusions}
\label{sec4}

In the present paper we have discussed the three-nucleon
bound system solution by combining the shell-model approach
with the Faddeev method. The use of Faddeev's decomposition
reduces the basis and allows to perform shell-model
calculations in significantly larger model spaces than
in the traditional shell-model approach. We were able to calculate
with the model spaces which included up to $32\hbar\Omega$
($N_{\rm max}=32$) harmonic-oscillator excitations.

We employed effective interactions, which take into account
the two-body correlations, in the calculations.
These effective interactions were derived in the 
two-particle relative coordinate channels from the 
Nijmegen II and Reid93 nucleon-nucleon potentials
and subsequently used in the three-body calculation.

As our results depend on the model-space size parameter 
$N_{\rm max}$ and on the harmonic-oscillator frequency
$\Omega$,
we tested the convergence in both these parameters.
Even for the largest model spaces we have not reached
complete convergence with respect to $\Omega$ in the whole
range of the used values. However, for $\hbar\Omega = 22-24$
MeV our results start to converge to the binding energies
obtained in the standard Faddeev calculations.   
As we include more partial waves in the nucleon-nucleon
potential, typically up to $j=6$,
our results seem to confirm the statements in Ref.
\cite{CPFG85} that the 34-channel standard Faddeev
calculation converged within 0.01 MeV. 
We have seen, in fact, that the higher partial waves of the
nucleon-nucleon interaction are not significant.
However, we believe a proper treatment of the metric 
operator (\ref{metric}) is important.

We observed that the convergence was rather slow with some 
dependence on the type of the nucleon-nucleon potential. 
In smaller model spaces we cannot
reproduce correctly the ground-state energy and the radius 
at the same time for any choice of $\Omega$. The wave function
probability distribution shows good stability as well as
the binding-energy difference between $^3$H and $^3$He.
Our results show, that an $8\hbar\Omega$ calculation accesible 
by the standard shell-model approach describes the ground states
properties within 10\% of exact values.
Modifications of effective interactions are
possible to improve the description in
small spaces. Examples of such modifications
can be found in Refs. \cite{ZBVHS,NB96}.

An unusual feature of the present approach
is the convergence from below. This is caused
by the use of effective interactions instead of the
free nucleon-nucleon interactions. The effective interactions
we employed were too strong. Obviously, it is possible to test
convergence properties of alternative effective interactions, 
as discussed, e.g., in Ref. \cite{HJKO95}.
It should be
noted, however, that the effective interaction should meet
the criterium $V_{\rm eff}\rightarrow V$ for 
$N_{\rm max}\rightarrow \infty$ in order to converge
to the exact solutions. 

The formalism discussed here may be generalized for more
complex systems as well. In particular, we are working on 
a generalization of the formalism for the $A=4$ system, which
relies on some results presented in this work. 
Also, it may be used to solve the three-nucleon system
bound in a harmonic-oscillator well. Then, from those solutions
three-body effective interactions can be constructed.
Such interactions, after transformation to an appropriate  
three-particle basis can serve as an input to standard
shell-model calculations for light nuclei.

\acknowledgements{
This work was supported by the NSF grant No. PHY93-21668.
P.N. also acknowledges partial support from 
the Czech Republic grant GA ASCR A1048504.
}

\begin{figure}
\caption{Ground-state energy, in MeV, dependence 
on the maximal number of harmonic-oscillator excitation allowed
in the model space for the Nijmegen II potential.
Results for $\hbar\Omega=14, 17, 19, 21, 22, 24$ MeV are presented.
The dotted line represents the result -7.62 MeV of the 34-channel Faddeev 
calculation reported in Ref. \protect{\cite{FPSS93}}.
}
\label{figniie}
\end{figure}

\begin{figure}
\caption{Ground-state energy, in MeV, dependence  
on the maximal number of harmonic-oscillator excitation allowed
in the model space for the Reid93 potential. 
Results for $\hbar\Omega=14, 17, 19, 22, 23, 24$ MeV are presented.
The dotted line represents the result -7.63 MeV of the 34-channel Faddeev 
calculation reported in Ref. \protect{\cite{FPSS93}}.
}
\label{figr93e}
\end{figure}

\begin{figure}
\caption{Point-nucleon radius, in fm, dependence
on the maximal number of harmonic-oscillator excitation allowed
in the model space for the Nijmegen II potential.
Results for $\hbar\Omega=14, 17, 19, 21, 22, 24$ MeV are presented.
}
\label{figniir}
\end{figure}

\begin{figure}
\caption{Point-nucleon radius, in fm, dependence
on the maximal number of harmonic-oscillator excitation allowed
in the model space for the Reid93 potential.
Results for $\hbar\Omega=14, 17, 19, 22, 23, 24$ MeV are presented.
}
\label{figr93r}
\end{figure}

\begin{figure}
\caption{$D$-state and $S'$-state probability, in \%, dependence
on the maximal number of harmonic-oscillator excitation allowed
in the model space for the Reid93 potential.
Results for $\hbar\Omega=19, 24$ MeV are presented.
}
\label{figwave}
\end{figure}

\begin{figure}
\caption{Dependence of the energy difference, in MeV, 
between the binding energies of $^3$H and $^3$He
on the maximal number of harmonic-oscillator excitation allowed
in the model space for the Reid93 potential.
Results for $\hbar\Omega=19, 24$ MeV are presented.
}
\label{figcoul}
\end{figure}

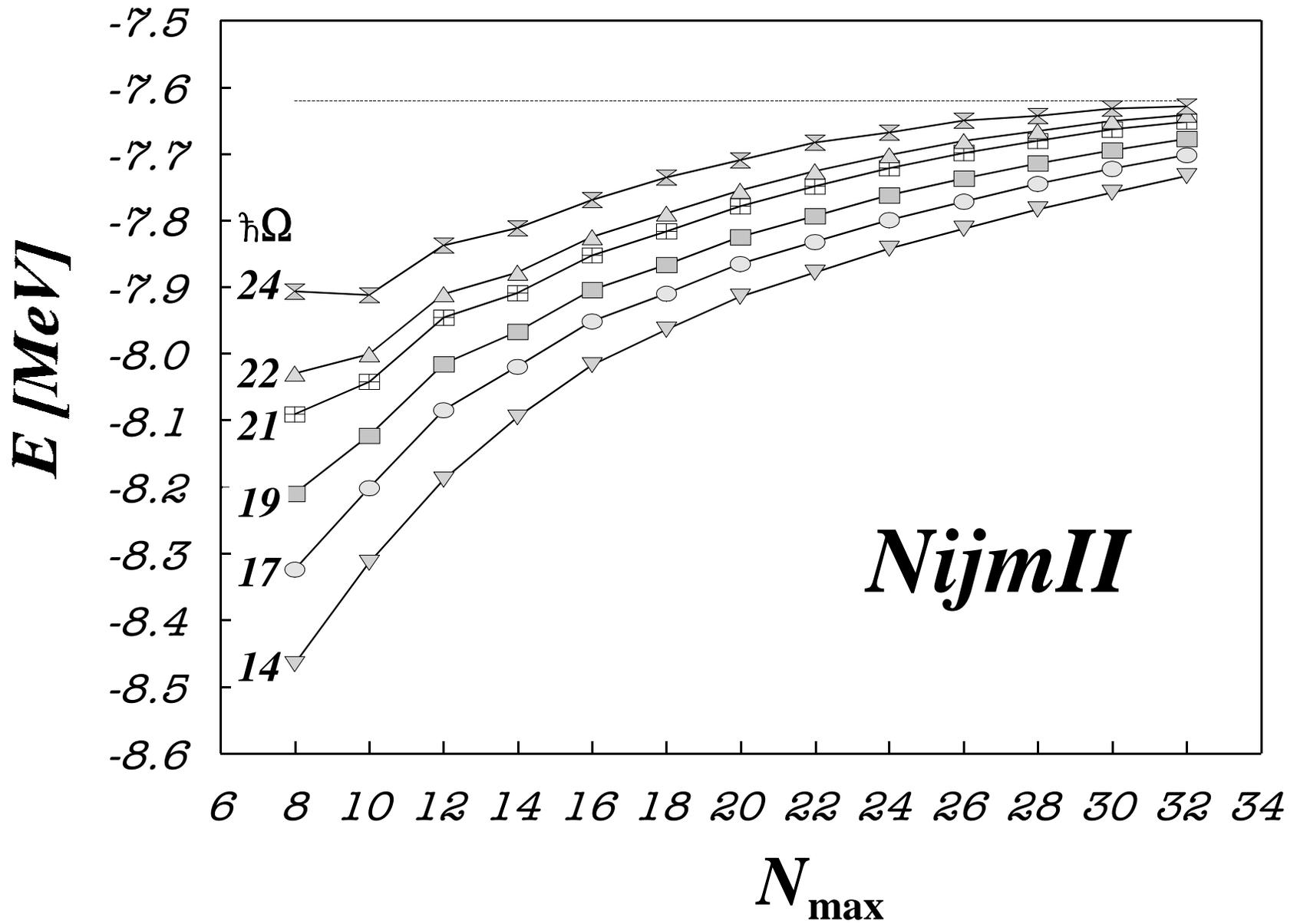
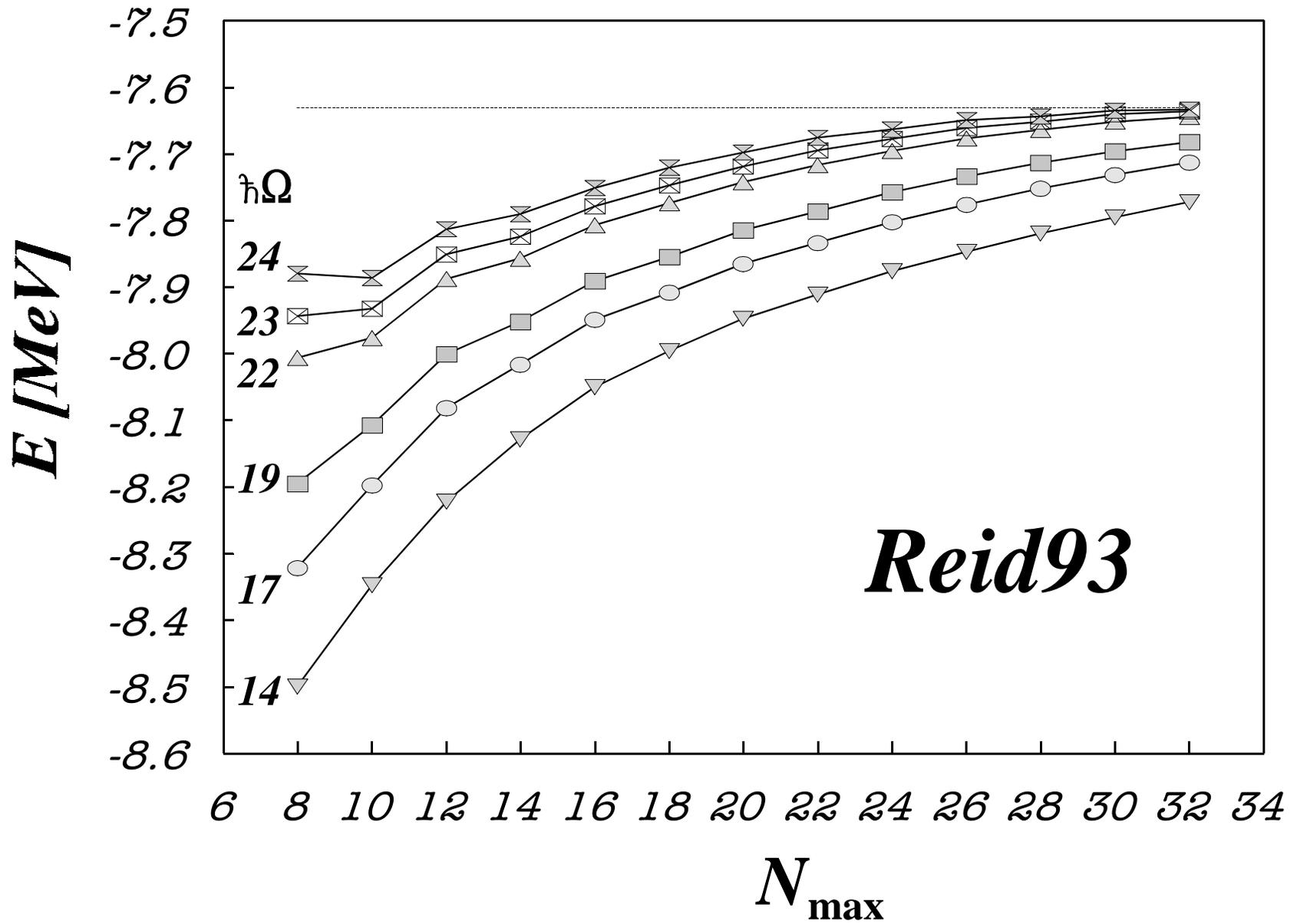
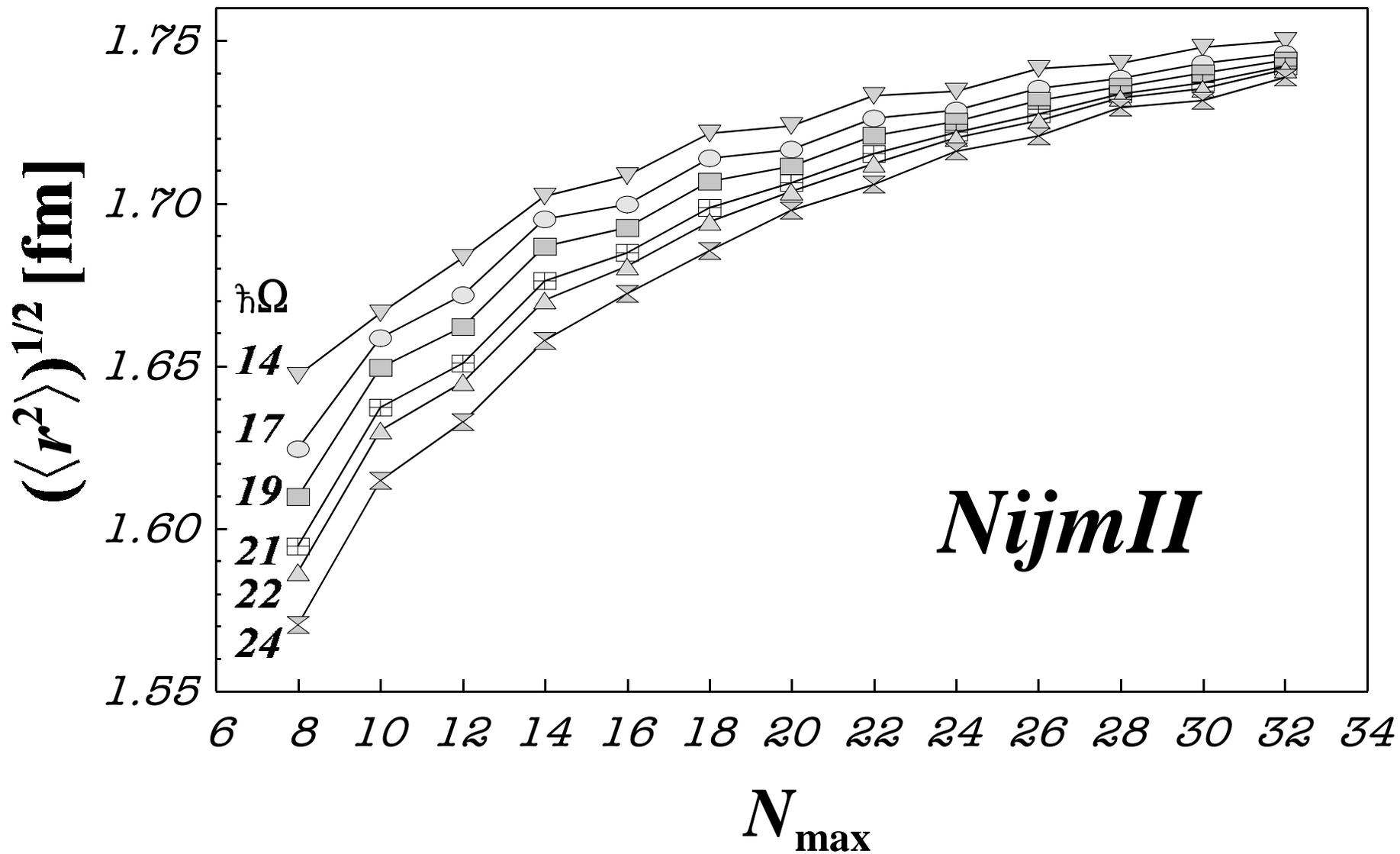
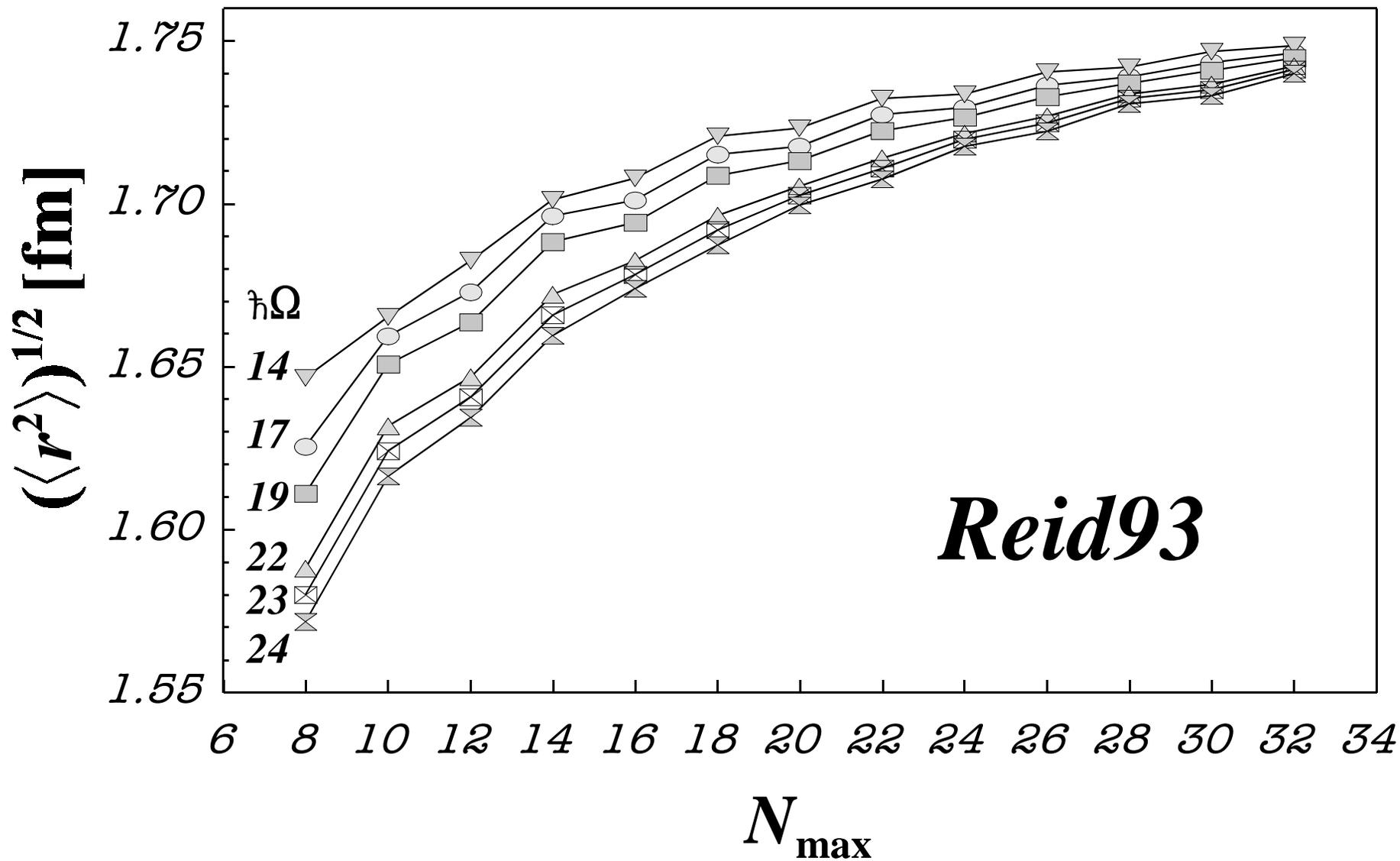
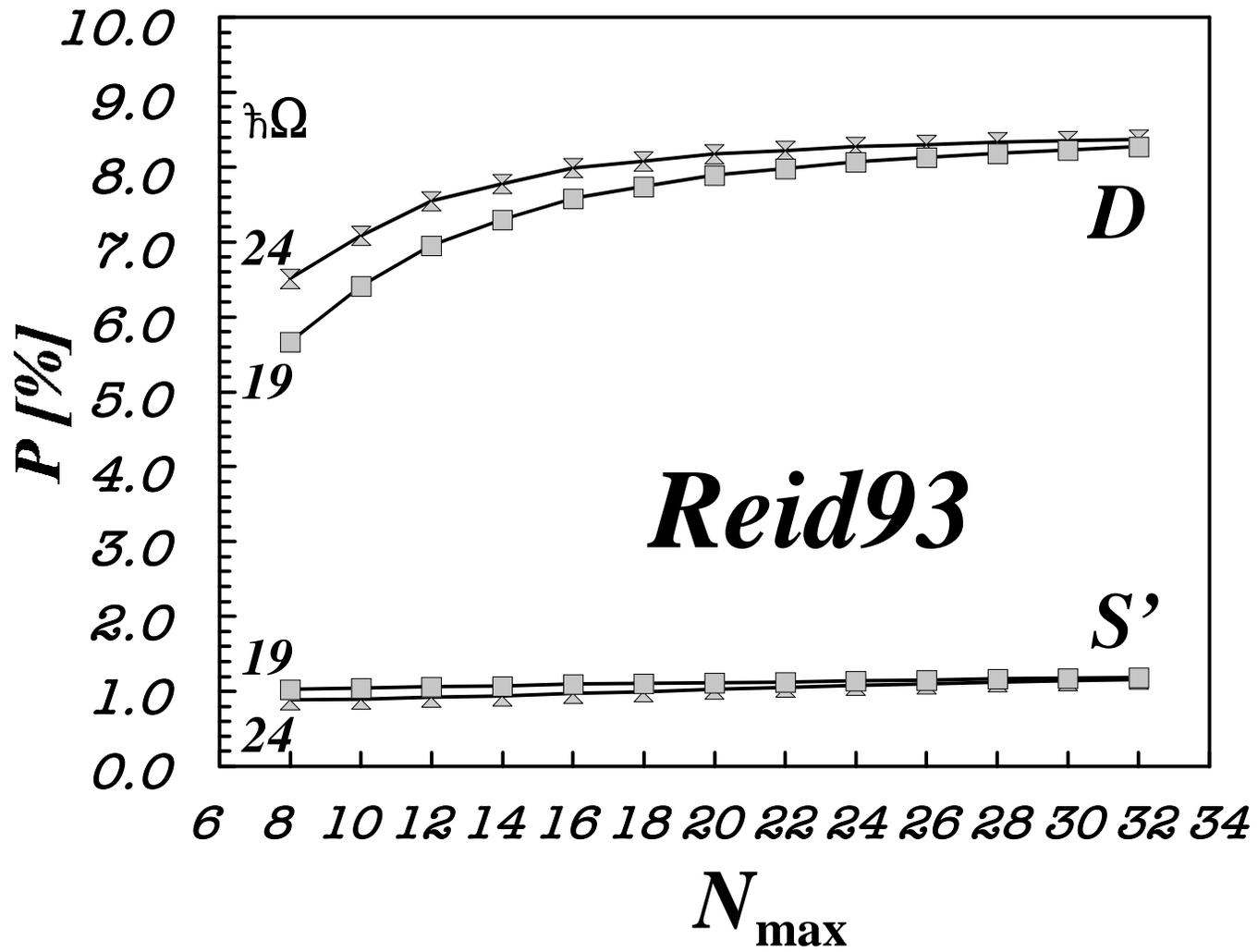
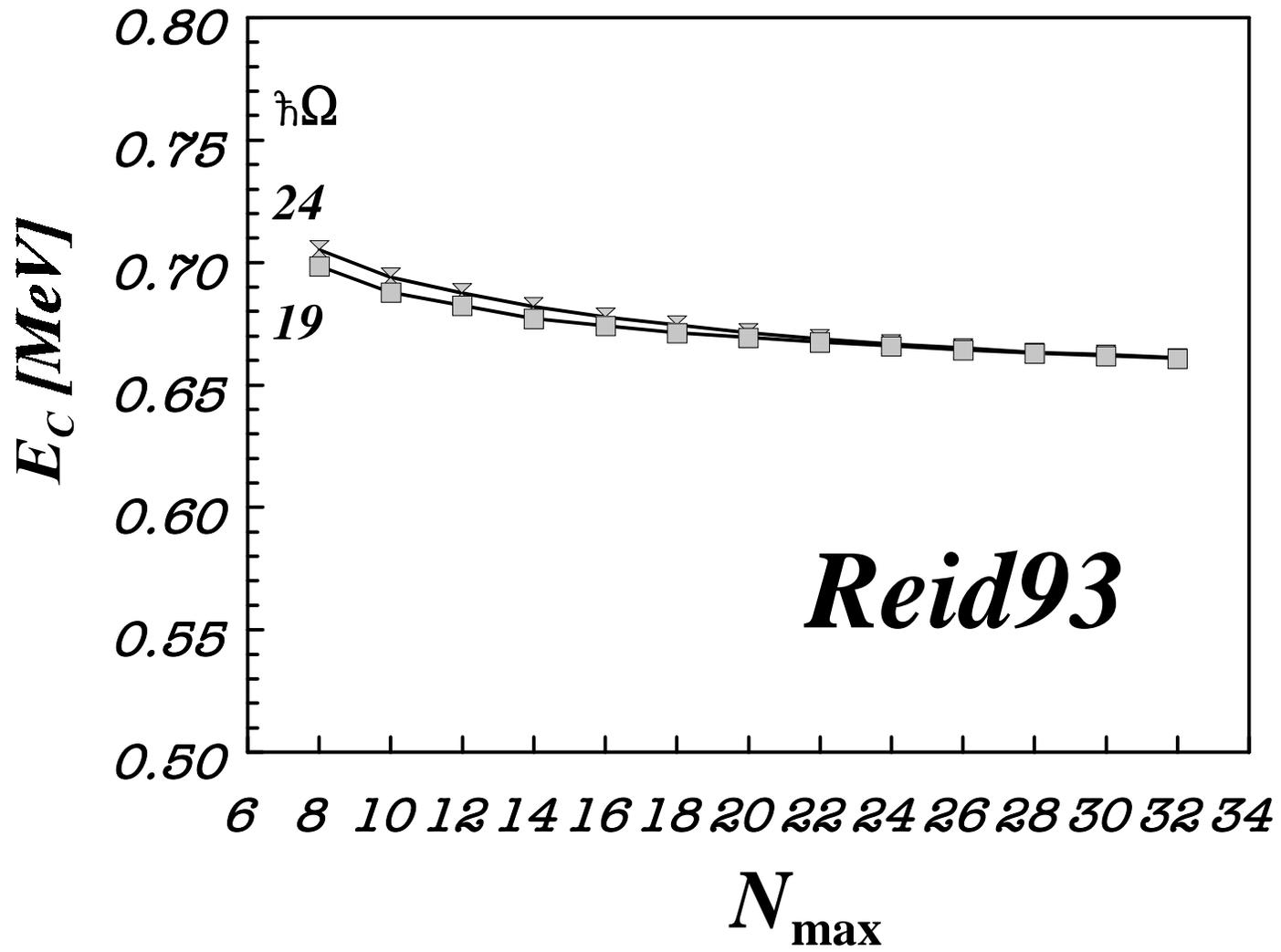
\end{document}